\documentclass[%
 aip,
 amsmath,amssymb,
twocolumn,10pt
]{revtex4-2}

\usepackage{graphicx}
\usepackage{dcolumn}
\usepackage{bm}
\usepackage[utf8]{inputenc}
\usepackage[T1]{fontenc}
\usepackage{mathptmx}
\usepackage{etoolbox}
\usepackage{xcolor}
\usepackage{ulem}
\usepackage{algpseudocode}

\makeatletter
\def\@email#1#2{%
 \endgroup
 \patchcmd{\titleblock@produce}
 {\frontmatter@RRAPformat}
 {\frontmatter@RRAPformat{\produce@RRAP{*#1\href{mailto:#2}{#2}}}\frontmatter@RRAPformat}
 {}{}
}
\makeatother
\begin{document}

\preprint{AIP/123-QED}

\title[Communication]{An efficient electrostatic embedding QM/MM method using periodic boundary conditions based on particle-mesh Ewald sums and electrostatic potential fitted charge operators}

\author{Simone Bonfrate}
\affiliation{Aix-Marseille Univ, CNRS, ICR, Marseille, France.}
\author{Nicolas Ferr\'e}
\affiliation{Aix-Marseille Univ, CNRS, ICR, Marseille, France.}
\author{Miquel Huix-Rotllant}
\email{miquel.huix-rotllant@cnrs.fr}
\affiliation{Aix-Marseille Univ, CNRS, ICR, Marseille, France.}

\date{\today}

\begin{abstract}
Hybrid quantum mechanics / molecular mechanics (QM/MM) models successfully describe the properties of biological macromolecules. However, most QM/MM methodologies are constrained to unrealistic gas phase models, thus limiting their applicability. In the literature, several works have attempted to define a QM/MM model in periodic boundary conditions (PBC) but frequently the models are too time-consuming for general applicability to biological systems in solution. Here, we define a simple and efficient electrostatic embedding QM/MM model in PBC combining the benefits of electrostatic potential fitted (ESPF) atomic charges and particle-mesh Ewald sums, that can efficiently treat systems of arbitrary size at a reasonable computational cost. To illustrate this, we apply our scheme to extract the lowest singlet excitation energies from a model for \textit{arabidopsis thaliana} cryptochrome 1 containing circa 93000 atoms, reproducing accurately the experimental absorption maximum.
\end{abstract}

\maketitle

Embedding methods in quantum chemistry allow reducing the overall computational cost by treating a small subsystem of atoms with an accurate theoretical method while treating the rest of the system in a cheaper and often less accurate approach.\cite{Jones2020} In such embedding schemes, the total energy is computed as the sum of energies of the constituent subsystems plus some interaction terms between each fragment.\cite{Fedorov2012} One of the most popular embedding methods for treating biological macromolecules is quantum mechanics / molecular mechanics (QM/MM),\cite{Cui2021} in which the energy is expressed as,
\begin{align}
 E=E_{QM}+E_{MM}+E^{int} \, ,
\label{eqn:st_qmmmm}
\end{align}
where $E_{QM}$ is the energy of the (small) QM subsystem, $E_{MM}$ is the energy of the (large) MM subsystem, and $E^{int}$ is the interaction term between them. Usually, the interaction is electrostatic, complemented with other pairwise atom-atom interactions. The majority of \textit{ab initio} QM/MM methods have been formulated employing an electrostatic Coulomb interaction between QM and MM subsystems, the complete macromolecular system being in the gas phase.\cite{HuixRotllant2021} There exist several attempts in the literature to formulate an \textit{ab initio} QM/MM method for models of macromolecules surrounded by an extended environment (solvent, membrane, etc.), either using non-periodic continuum models\cite{Falbo2022} or periodic boundary conditions (PBC) employing the Ewald summation technique.\cite{Nam2005,Riccardi2005,Laino2005,Laino2006,Seabra2007,Walker2008,SanzNavarro2011,Holden2013,Nam2014,Holden2015,Vasilevskaya2016,Giese2016,Nishizawa2016,Kawashima2019,Holden2019,Pederson2022} Most QM/MM PBC formulations rely on atomic point charges for efficiently representing the long-range QM-QM interactions such as Mulliken, \cite{Nam2005,Riccardi2005,Seabra2007,Walker2008,Nishizawa2016,Nam2014}, ChElPG, \cite{Holden2013,Holden2015,Holden2019}, ESP,\cite{Vasilevskaya2016} or other types.\cite{Kawashima2019,Pederson2022} Such methods mainly use Ewald pair potentials, \cite{Nam2005,Holden2013,Holden2015,Holden2019,Vasilevskaya2016}, standard Ewald, \cite{Riccardi2005,SanzNavarro2011,Kawashima2019,Laino2006} or exploiting the efficiency of particle-mesh Ewald (PME) method.\cite{Darden1993,Essmann1995,Sagui2004} The PME, which is state-of-the-art algorithm for efficiently calculating long-range interactions in large MM systems, has mainly been implemented for semi-empirical\cite{Walker2008,Seabra2007,Nishizawa2016,Nam2014} QM/MM methods. We are aware of only two recent articles reporting its use in \textit{ab initio} QM/MM methods\cite{Giese2016,Pederson2022}.

\begin{figure}
 \centering
 \includegraphics[width=0.30\textwidth,keepaspectratio]{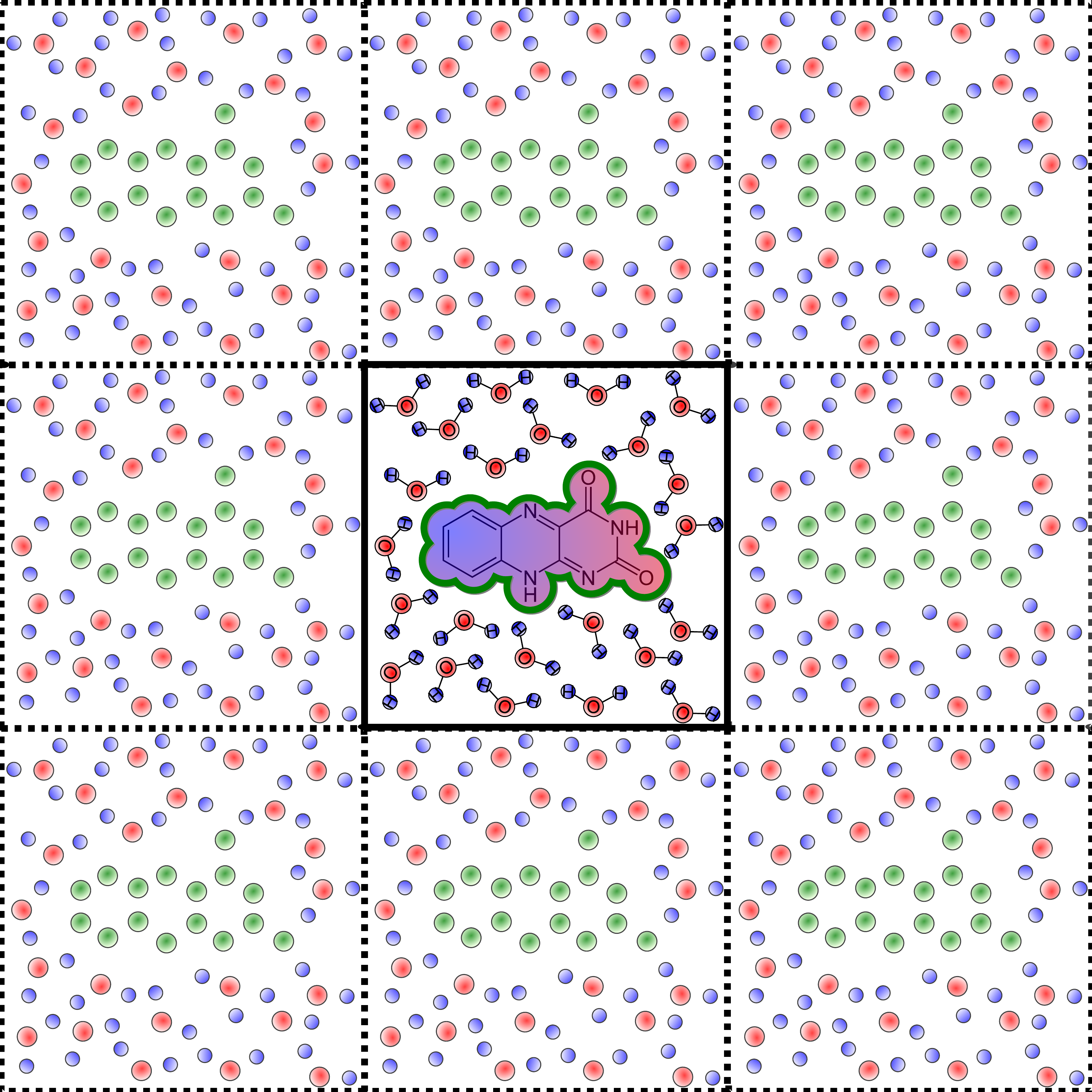}
 \caption{Schematic representation of the electrostatic embedding QM/MM PBC method described here for a chromophore (QM) in water (MM). In the original cell (center), the QM subsystem is represented by a quantum charge density. In the replica cells, the QM atoms are represented as ESPF point charges (green circles). MM atoms are represented with blue and red point charges. All point charges are used to polarize the QM density in the original cell.}
 \label{fig1}
\end{figure}

Here, we combine the advantages of electrostatic potential fitted (ESPF) charges,\cite{Ferre2002,HuixRotllant2021} and PME potentials to formulate an efficient \textit{ab initio} electrostatic embedding QM/MM PBC method, defining a unified consistent embedding energy from an interaction hamiltonian. Our formulation takes full computational advantage of PME, it reduces the number of integrals to be computed and can be applied to any \textit{ab initio} self-consistent field method. The definition of ESPF QM/MM PBC interaction (see Fig.\ \ref{fig1}) is based on the pairwise electrostatic interaction energy between QM and MM subsystems defined in terms of potentials,
\begin{align}
 \label{eq:intener}
 E^{int}=\sum_A^{N_{QM}}{q_A\Phi^{MM}_A} + \frac{1}{2}\sum_A^{N_{QM}}{q_A\Phi^{QM}_A} \, ,
\end{align}
where the first term accounts for the QM-MM interactions and the second term for the QM-QM interactions, with the $1/2$ factor to avoid double counting. Hereafter, we use the common index notation upper and lower case letters to indicate QM and MM atoms respectively. The $N_{QM}$ is the number of QM atoms, $q_A=Z_A-Q_A$ are their partial charges, defined as the difference between the atomic charge $Z_A$ and the electronic charge population $Q_A=\sum_{\mu\nu}{P_{\mu\nu}Q_{A,\mu\nu}}$, obtained as the contraction between the quantum density matrix $\mathbf{P}$ and an atomic charge operator matrix $\mathbf{Q}_A$ to be defined later on.  The notation $\Phi_A$ is a short-hand notation for the external potential felt at atom position A, that is, $\Phi_A=\Phi({\bf r}_A)$. 

It must be immediately emphasized that $\partial\Phi^{MM}_A/\partial P_{\mu\nu}=0$ while $\partial\Phi^{QM}_A/\partial P_{\mu\nu}\ne 0$. Taking this into account, we can obtain an interaction operator by deriving the interaction energy with respect to any density matrix element $P_{\mu\nu}$, leading to
\begin{align}
 h^{int}_{\mu\nu}=-\sum_A^{N_{QM}}{Q_{A,\mu\nu}\left(\Phi^{MM}_A+\Phi_A^{QM}\right)}=-\sum_A^{N_{QM}}{Q_{A,\mu\nu}\Phi_A}
 \, , 
 \label{eq:intop}
\end{align}
where we defined the total potential $\Phi_A=\Phi_A^{MM}+\Phi_A^{QM}$.

Up to this point, this is a general formulation for a QM/MM embedding when using charge operators. The different QM/MM models are then distinguished by defining the MM energy and the external electrostatic potential. For example, in the straightforward QM/MM implementations of pairwise Coulomb interactions without PBC, the QM/MM procedure is simple by defining the external electrostatic potential as
\begin{align}
 \Phi_A=\Phi^{MM}_{A} &= \sum_{j=1}^{N_{MM}} \frac{q_j}{\left| \mathbf{r}_{Aj} \right|} \, .
\end{align}
Here, $\mathbf{r}_{Aj}=\mathbf{r}_{A}-\mathbf{r}_{j}$ is the distance vector between the two charges. In this case, $\Phi_A^{QM}=0$ and therefore the interaction energy and operator matrix elements can be computed with the sole knowledge of the potential generated by MM atoms on QM centers.

The use of PBC allows to account for the long-range interactions and to build a more realistic model of the macromolecule interactions with the solvent. However, the introduction of Coulomb interaction with replicas results in slow and conditionally convergent interaction energy. The employment of Ewald summation technique to reach a faster convergence by introducing a range-separated electrostatic interaction,\cite{Ewald1921} allows to split the pairwise Coulombic interaction energy for a system containing $N$ point charges into three contributions,
\begin{align}
 E &= E^{short} + E^{long} + E^{self} \, ,
\end{align}
in which we define short-range energy as
\begin{align}
 E^{short} &= \frac{1}{2} \sum_{\mathbf{n}=\mathbf{0}}{\sideset{}{'}\sum_{\alpha,\beta =1}^{N}{\frac{q_{\alpha}q_{\beta}}{\left| \mathbf{r}_{\alpha\beta \mathbf{n}} \right|} \mathrm{erfc}(\beta \left| \mathbf{r}_{\alpha\beta \mathbf{n}} \right|)}} \, ,\label{eqn:ew_short}
\end{align}
the long-range energy as
\begin{align}
 E^{long} &= \frac{1}{2 \pi V}\sum_{\mathbf{m} \neq \mathbf{0}}\frac{e^{-\frac{\pi^2\mathbf{m}^2}{\beta^2}}}{\mathbf{m}^2} \left| S(\mathbf{m}) \right| ^2 \, , \label{eqn:ew_long} 
\end{align}
and the self-interaction energy as
\begin{align}
 E^{self} &= -\frac{\beta}{\sqrt{\pi}}\sum_{\alpha=1}^{N}q_{\alpha}^2 \, . \label{eqn:ew_self}
\end{align}
In these formulas, ${\bf r}_{\alpha\beta\mathbf{n}}={\bf r}_\alpha-{\bf r}_{\beta}+\mathbf{n}L$, where $L$ is the length of the unitary cell (cubic) box and $\beta$ is the range separation parameter that controls the rates of converging of the first two terms. $E^{short}$, which recovers the short-range part of the interactions, is computed in real space and contains a summation over the original box and all the replicas which are described by the $\mathbf{n}$ vectors. The prime in the second sum of Eq.\ \ref{eqn:ew_short} means that we are excluding those terms for which $\alpha=\beta$ when $\mathbf{n} = \mathbf{0}$, while $\mathrm{erfc}(x)$ is the complementary error function which is defined as $\mathrm{erfc}(x)=1-\mathrm{erf}(x)$, where $\mathrm{erf}(x)$ is the error function. $E^{long}$, including the long-range part of the interactions, is computed in the reciprocal space, where the summation runs over all the reciprocal space vectors $\mathbf{m}$. The so-called structure factors are defined by $S(\mathbf{m}) = S_{QM}(\mathbf{m})+S_{MM}(\mathbf{m})$, and
\begin{align}
\label{eq:struc_fact}
 S_{QM}(\mathbf{m}) = \sum_{A=1}^{N_{QM}} q_{A} e^{2\pi i \mathbf{m} \cdot \mathbf{r}_{A}} \ \  ; \ \ 
 S_{MM}(\mathbf{m}) = \sum_{i=1}^{N_{MM}} q_{i} e^{2\pi i \mathbf{m} \cdot \mathbf{r}_{i}} \, . 
\end{align}
Note that, within the long-range energy expression, the interactions in the original cell for which $\alpha=\beta$ are not omitted from the summation and, instead, the contribution to the energy due to such terms is removed by the introduction of self-interaction energy ($E^{self}$) which is a constant correction term.

For the QM/MM model in PBC, it is necessary to define new expressions for MM energy and $\Phi_A$ out of the Ewald energy expression. The purely MM electrostatic energy contribution can be easily obtained from Eqs.\ \ref{eqn:ew_short} to \ref{eqn:ew_self} by restricting the summations to MM atoms only. The electrostatic potentials in QM/MM PBC procedure have both MM and QM parts. The MM potential contains short and long-range contributions, 
\begin{align}
 \Phi^{MM}_A &= \Phi_A^{short,MM} + \Phi_A^{long,MM} \, ,
\end{align}
while the QM potential contains in addition to the short and long-range potentials, a self-interaction and correction terms,
\begin{align}
 \Phi^{QM}_A &= \Phi_A^{short,QM} + \Phi_A^{long,QM} + \Phi_A^{self,QM} + \Phi_A^{corr,QM} \, .
\end{align}
Here, $\Phi_A$ has to be understood as the electrostatic potential calculated at the position of QM atom $A$ in the original cell, that is, $\Phi_A=\Phi(\mathbf{r}_{A}+\mathbf{0}L)$. The expressions for the short-range MM and QM potentials are given by
\begin{alignat}{2}
 \Phi_{A}^{short,MM} &= \hspace{0.5em}
 &&\sum_{\mathbf{n}=\mathbf{0}}\sum_{i=1}^{N_{MM}} \frac{q_{i}}{\left| \mathbf{r}_{Ai \mathbf{n}} \right|} \mathrm{erfc} \left( \beta \left| \mathbf{r}_{Ai \mathbf{n}} \right| \right) \nonumber \\
 \Phi_{A}^{short,QM} &= \hspace{0.5em} &&\sum_{\mathbf{n}\neq\mathbf{0}}\sum_{B=1}^{N_{QM}} \frac{q_{B}}{\left| \mathbf{r}_{AB \mathbf{n}} \right|} \mathrm{erfc} \left( \beta \left| \mathbf{r}_{AB \mathbf{n}} \right| \right) \, , 
\end{alignat}
and the long-range MM and QM potentials are given by
\begin{align}
\label{eqn:long_expot}
 \Phi_{A}^{long,MM} &= \frac{1}{\pi V} \sum_{\mathbf{m}\neq \mathbf{0}} \frac{e^{\frac{-\pi^2 \mathbf{m}^2}{\beta^2}}}{\mathbf{m}^2} \Re \left[ e^{-2\pi i \mathbf{m} \cdot \mathbf{r}_A} S_{MM}(\mathbf{m}) \right] \nonumber \\
 \Phi_{A}^{long,QM} &= \frac{1}{\pi V} \sum_{\mathbf{m}\neq \mathbf{0}} \frac{e^{\frac{-\pi^2 \mathbf{m}^2}{\beta^2}}}{\mathbf{m}^2} e^{-2\pi i \mathbf{m} \cdot \mathbf{r}_A} S_{QM}(\mathbf{m}) \, , 
 \end{align}
where $\Re (z)$ is the real part of $z$.
For constructing the total QM potential, two extra terms have to be considered, the self-interaction ($\Phi_A^{self,QM}$) and the correction potentials ($\Phi_A^{corr,QM}$). The self-interaction potential is defined as
\begin{align}
 \Phi_{A}^{self,QM} &= - \frac{2\beta}{\sqrt{\pi}}q_{A} \, ,
\end{align}
while the correction potential is defined as
\begin{align}
\label{eqn:corr_pot}
 \Phi_{A}^{corr,QM} &= -\sum_{B=1}^{N_{QM}} \frac{q_{B}}{\left| \mathbf{r}_{AB\mathbf{0}} \right|} \mathrm{erf} \left( \beta \left| \mathbf{r}_{AB\mathbf{0}} \right| \right) \, .
\end{align}
These two terms arise from spurious interactions that need to be removed.\cite{Nam2005} In this framework, extra energy terms are easily included as extra potential sources (see for example the surface-dipole and the non-neutral cell correction terms in the supporting information).

The ESPF method ensures the uniqueness of the resulting QM/MM potential energy surface, an absolute requirement for obtaining accurate molecular gradients at a reasonable computational cost.\cite{Melaccio2011} Therefore, ESPF guarantees that the correction term in eq.\ \ref{eqn:corr_pot} is equivalent to the interaction term that should be removed from eq.\ \ref{eqn:long_expot}. In the ESPF procedure, the charge operator matrix elements are fitted to QM-only electrostatic integrals computed on a numerical grid constructed around the molecule.\cite{Ferre2002} For obtaining the charge operators used in eq.\ \ref{eq:intop}, a system of equations,
\begin{align}
 \sum_A\frac{Q_{A,\mu\nu}}{\left|{\bf r}_k-{\bf r}_A\right|}=\int{d^3{\bf r}\;\chi_\mu^*({\bf r})\frac{1}{|{\bf r}-{\bf r}_k|}\chi_\nu({\bf r})} \, ,
\end{align}
 has to be solved. The ${\bf r}_k$ are the point coordinates of a Lebedev atom-centered grid defined around the molecule, and $\chi$ are the atomic orbitals. A correction is added to the charge operator matrix elements to ensure the conservation of the total charge of the QM subsystem.\cite{HuixRotllant2021}
 
While Ewald summation has been undoubtedly useful for handling long-range interactions inside the PBC framework, its original formulation scales like $\bm{O}(N^2)$ and becomes soon computationally unfeasible when the MM system is large. As a consequence, different approaches, aimed at reducing the algorithmic complexity have been proposed. \cite{Toukmaji1996,Wheeler2002,Shan2005,Nestler2015} The model described in this work makes use of the Smooth Particle Mesh Ewald (SPME) method, firstly introduced by Pedersen and coworkers,\cite{Darden1993,Essmann1995,Sagui2004} which features a reduced complexity of $\bm{O}(N\log(N))$.
The main idea behind PME consists in approximating the structure factor (eq. \ref{eq:struc_fact}) by interpolating the complex exponential. While the original Particle Mesh Ewald method makes use of Lagrangian interpolation, the SPME is based on cardinal B-spline interpolation, which allows analytic differentiation. Within SPME, the long-range potential could be expressed as 
\begin{align}
 \Phi_{A}^{long} \approx \sum_{ \mathbf{k}} \theta_p(u_{1j} - k_1) \theta_p(u_{2j} - k_2) \theta_p(u_{3j} - k_3) 
 \left(
 G\star Q
 \right)
 ( \mathbf{k})
\end{align}
where $\theta_p$ is the $p$-th order cardinal B-spline functions, $\mathbf{u}$ contains the scaled fractional coordinates of a point of coordinates $\mathbf{r}$ in the original cell, $\mathbf{n}$ are the real-space grid points vectors, $G\star Q$ is the convolution of $G$, the generalized influence function defined by its Fourier transform, and $Q$, the generalized grid multipolar array.\cite{Sagui2004}

\begin{figure}
 \centering
 \includegraphics[width=0.5\textwidth,keepaspectratio]{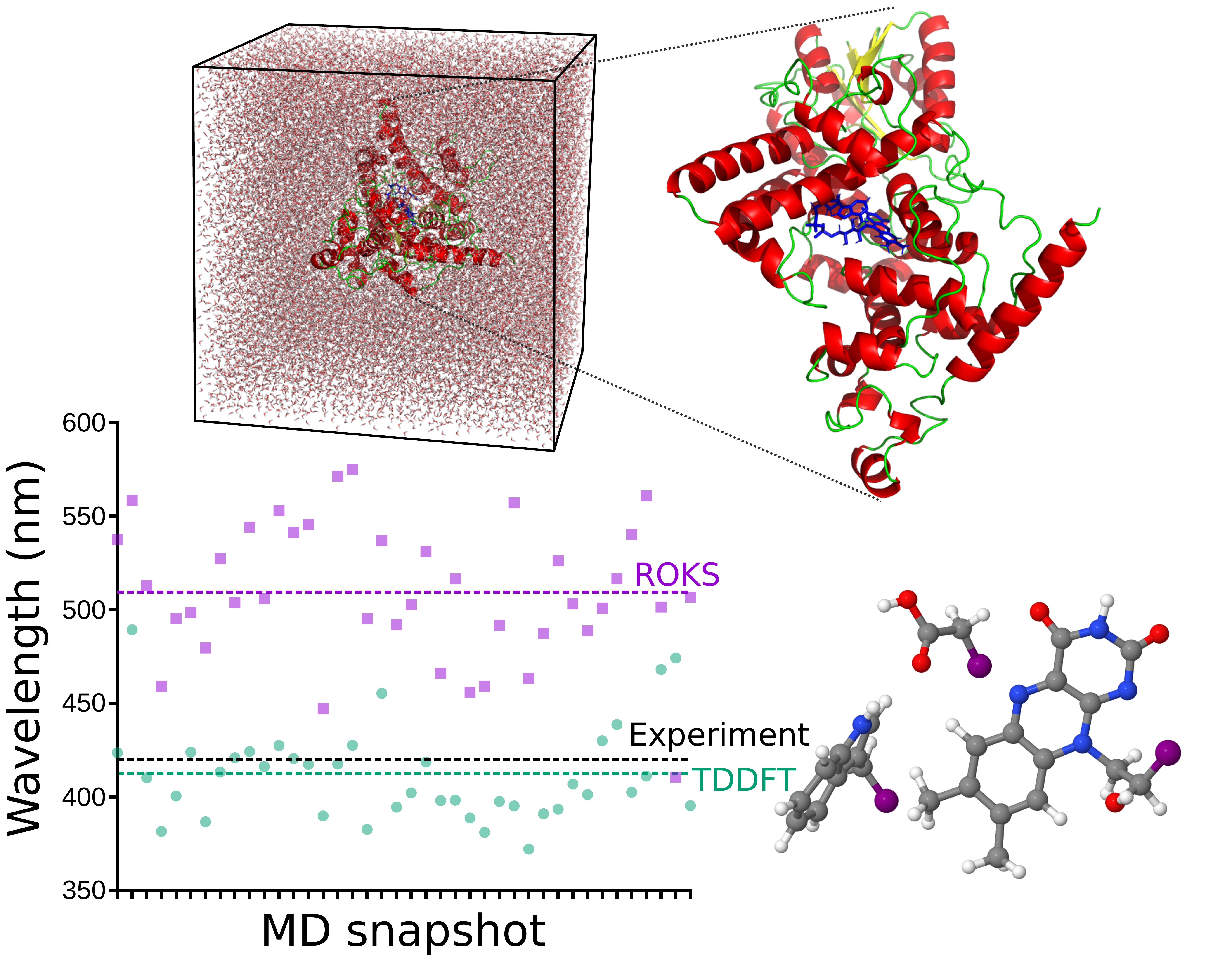}
 \caption{(top) ESPF QM/MM PBC model of \textit{arabidopsis thaliana} cryptochrome 1 in a water box. The isoalloxazine, D396 and W400 (64 atoms in total) are treated at the QM level, with the link atoms shown in purple; (bottom, left) Excitation energies for several snapshots computed with restricted open-shell Kohn-Sham and time-dependent density functional theory. The mean values (dashed lines) are compared to experimental values from Ref.\ \citenum{Ahmad2002}.}
 \label{fig2}
\end{figure}

The presented methodology, combining PME and ESPF charges, represents a consistent formulation between the energy and the Hamiltonian and an efficient method for computing the ground and excited state energies. To test this, we extract the lowest singlet excitation energies of 40 snapshots of \textit{arabidopsis thaliana} cryptochrome 1 (see Fig.\ \ref{fig2} and the supporting iformation for the computational details). On the one hand, time-dependent density functional theory (TDDFT) has been performed on top of the QM/MM PBC ground state Kohn-Sham reference. In this model, we consider that the external potential is fixed in the excited state calculations, and therefore, the only excitation process occurs in the original cell, while the replicas remain in the ground state. No response terms have been added to the TDDFT equations. This approximate model, for which the average excitation energy is around 412 nm $\pm$ 26 nm, is in excellent accordance with the experimental absorption maximum of 420 nm.\cite{Ahmad2002} The blue-shift can be attributed to the lack of vibronic effects, which are known to be important in the absorption spectra of isoalloxazine.\cite{Schwinn2020} On the other hand, we have implemented a restricted open-shell Kohn-Sham (ROKS) model to extract the lowest energy excited state directly from the SCF.\cite{Kowalczyk2013} In this model, both the chromophore in the original cell and the replicas are excited. In this case, the average excitation energy is 518 nm $\pm$ 35 nm, thus underestimated by around 100 nm with respect to experiments. Of course, this model is limited by the fact that excited states are represented by a single configuration, but also the probably unrealistic situation that the photoexcited protein is surrounded by simultaneously excited proteins in the replicas.

In conclusion, we have presented an efficient QM/MM formulation in periodic boundary conditions based on electrostatic potential fitted charges and smooth particle-mesh Ewald sums. The method scales approximately like $O(\alpha\cdot N_{MM}^{1.3})$ ($\alpha=4\cdot 10^{-6}$, see supporting information for further details), opening up the route for a general application of QM/MD simulations in large-sized periodic systems. This will require the computation of analytic energy first derivatives, which we plan to develop in the future.

\begin{acknowledgements}
We acknowledge the support from ``Agence Nationle de la Recherche'' through the project MAPPLE (ANR-22-CE29-0014-01). Centre de Calcul Intensif d'Aix-Marseille is acknowledged for granting access to its high performance computing resources.
\end{acknowledgements}

\section*{Data availability statement}

The data that support the findings of this study are available from the corresponding author upon reasonable request.
 
\bibliography{refs}

\end{document}